\documentclass[fleqn,usenatbib]{mnras}

\usepackage{newtxtext,newtxmath}
\usepackage{longtable}
\usepackage[T1]{fontenc}
\usepackage[utf8]{inputenc}
\DeclareRobustCommand{\VAN}[3]{#2}
\let\VANthebibliography\thebibliography
\def\thebibliography{\DeclareRobustCommand{\VAN}[3]{##3}\VANthebibliography}

\usepackage{graphicx}	
\usepackage{amsmath}	
\newcommand{\minus}{\scalebox{0.75}[1.0]{$-$}}
\usepackage{subcaption} 
\usepackage{threeparttable} 
\usepackage{booktabs, caption, makecell} 
\usepackage{multirow} 
\usepackage{ragged2e} 
\usepackage{url} 
\usepackage{hyperref} 
\DeclareUnicodeCharacter{00A0}{~}
\DeclareUnicodeCharacter{2212}{-} 

\title[ULXs in NGC~5813]{Ultraluminous X-ray sources in the group-centric elliptical galaxy  NGC~5813}
\author[Rajalakshmi et al.]{T. R. Rajalakshmi, $^{1}$\thanks{E-mail: rajalakshmi.thevalil@gmail.com}
Somak Raychaudhury,$^{2,3,4}$
Indulekha Kavila,$^{1,3}$
and Gulab C. Dewangan$^{3}$
\\
$^{1}$ School of Pure and Applied Physics, Mahatma Gandhi University, Kerala 686560, India\\
$^{2}$ Department of Physics, Ashoka University, Sonepat, Haryana 131029, India\\
$^{3}$ Inter-University Centre for Astronomy and Astrophysics, Post Bag 4, Ganeshkhind, Pune 411007, India\\
$^{4}$ School of Physics and Astronomy, University of Birmingham, Birmingham B15~2TT, UK
}
\date{Accepted XXX. Received YYY; in original form ZZZ}
\pubyear{2024}

\begin{document}
\label{firstpage}
\pagerange{\pageref{firstpage}--\pageref{lastpage}}
\maketitle

\begin{abstract}
The number of Ultraluminous X-ray Sources (ULXs) is observed to be correlated with the current star formation rate in late-type galaxies and with the stellar mass in early-type galaxies (ETGs). Since there is very little gas, dust or star formation in ETGs, it has been suggested that most of the ULXs  associated with them could be high luminosity Low Mass X-ray Binaries (LMXBs) or foreground/background sources.
It has been reported that NGC~5813, the central dominant (cD) galaxy in the NGC~5846 group of galaxies, which shows signs of a possible recent merger event, has an unusually high number of ULXs.  We have undertaken a multi-epoch spectral study of the persistent ULXs in the galaxy using \textit{Chandra} and XMM-Newton observations.  Of the eight ULXs reported elsewhere, four have been re-identified, two are not consistently detected across all nine \textit{Chandra} observations, and two are found to be foreground sources.  One new persistent ULX has been identified.  We present a spectral analysis of the five ULXs with luminosity consistently greater than $10^{39}$ erg/s in nine \textit{Chandra}-ACIS observations, and assess their variability, adding data from XMM-Newton.  The association of these ULXs with globular clusters was examined: we find one ULX lying within the field of an HST observation within 0.1$^\prime$ of the centre of a globular cluster.  Optical and UV counterparts are found for another ULX.   One of the ULXs is found to be variable over the time scale of days, but there is no unambiguous evidence of longer-term variability for the remaining ULXs.
\end{abstract}




\section{Introduction}
Ultraluminous X-ray sources (ULXs) are extragalactic non-nuclear point-like sources whose X-ray luminosity is intermediate between the Eddington limit of accreting stellar mass black holes and supermassive black holes.  They have been proposed to host neutron stars or stellar-mass black holes, accreting at super-Eddington rates (with a not negligible geometrical beaming) or as Intermediate Mass Black Holes (IMBHs) accreting at sub-Eddington rates \citep[e.g.][]{FengSoria2011, KaaretFeng2017,King2023}. \par 
There are persistent as well as transient ULXs.  Persistent ULXs have luminosities above 10$^{39}$ erg/s in all observations, but transient ULXs overpass this luminosity only for limited periods, the estimated recurrence time varying from a few days to a few years \citep[e.g.][]{Strickland2001, Kong2005, vanHaaften2019, Salvagio2022, Brightman2023}.  A number of ULXs showing coherent pulsations, and therefore hosting without a doubt a neutron star as the compact object, have been discovered; these are known as ULX Pulsars or Pulsating ULXs (PULXs) \citep{Bachetti2014,Furst2016,Israel2017a,Israel2017b,Carpano2018,Satyaprakash2019,Rodriguez2020}. A number of candidate Neutron Star (NS) ULXs  have also been discovered; however, pulsations have not been confirmed in some of them \citep[e.g.][]{Pintore2017,Koliopanos2017, Walton2018, Maitra2019,Gurpide2021b,Amato2023}.\par

ULXs were initially detected in late-type galaxies and their numbers were found to be correlated with the star formation rate (SFR) \citep[e.g.][]{Gilfanov2004} of the galaxy.  Subsequent studies \citep[e.g.][]{Mineo2013}, have explored the correlation between the location of the ULX in a galaxy and the local SFR determined using \textit{GALEX} FUV and \textit{Spitzer} 24$\mu$m data. Although ULXs were thought to be rare in early-type galaxies (ETGs), as ellipticals have comparatively lower SFR than late-type galaxies, it became obvious that there are significant ULX populations in ETGs as well \citep[e.g.][]{Raychaudhury2008}. Some of them were associated with the globular clusters (GCs) of the galaxy \citep{Angelini2001}.  \citet{Dage2021} have identified 20 ULXs associated with GCs in ETGs within 20 Mpc.  \citet{Thygesen2023} has identified nine additional ULXs associated with GCs in ETGs within a distance of 70 Mpc.  Since the SFR in elliptical galaxies is expected to be quite low, it is possible that ULXs associated with them are Low Mass X-Ray Binaries (LMXBs) \citep{King2002}.  The number of ULXs in a galaxy is seen to be anti-correlated with the metallicity of the galaxy \citep[e.g.][]{Mapelli2009}.  In the case of ETGs, there is a nonlinear dependence on the stellar mass of the galaxy  \citep{Kovlakas2020}.   \par
A stellar-mass compact object accreting at super-Eddington accretion rates is likely to have a thick accretion disk and violent winds can be driven from inner regions \citep{Shakura1973,Poutanen2007}.  In general, ULX spectra can be fitted with two thermal components, a soft one and a hard one \citep{Stobbart2006,Pintore2015,Pinto2021, Gurpide2021, Amato2023}, with an additional non-thermal component \citep{Walton2018} when high quality NuSTAR observations are available (>10.0 keV).  The soft emission originates in the wind and the hard emission from the inner, distorted disk.  A quasi-thermal, multicolour disk blackbody is used to account for emission from the wind and a thermally Comptonized continuum to describe the emission from the inner disk and its down-scattered component \citep[e.g.][]{Middleton2015}.  The hard non-thermal component is due to the inverse compton scattering of disk photons by hot electrons in the Comptonizing corona (hybrid corona) of BH ULXs \citep{Gladstone2009b} or the accretion column in NS ULXs \citep{Walton2018}.   
 \citet{Sutton2013} identified three different spectral states of ULXs, which can be named as "broadened disc", characterised by a broad thermal component and expected to be found at (or slightly above) the Eddington rate, "hard ultraluminous" which shows a two-component spectrum, with the hard one dominating the emission, and accretion is expected to be significantly above Eddington, and a third state "soft ultraluminous" with the highest accretion rate and emission peaking at the softer end of the 0.3--10.0 keV range. It should be noted that the spectral appearance can be driven by the inclination angle of the systems with respect to our line of sight \citep{Middleton2015}.

In the initial, short-exposure XMM Newton and Chandra observations of ULXs, two-component models with a thermal component and a non-thermal component (an accretion disk with a powerlaw and a blackbody with a powerlaw component) were often used to describe the spectra \citep[e.g.][]{Miller2003, Cropper2004}. For low-quality ULX spectra, single-component models like multi-colour disk black body (MCD) or powerlaw models yield good fits \citep{Robert2004, Burke2013a, Burke2013b, Jithesh2014}.\par 
ULX spectra can vary with time. Spectral variability of ULXs whose spectra are well fitted with an absorbed powerlaw model, can be explored further by studying the correlation between $\Gamma$ and luminosity and, the correlation between inner disk temperature and disk luminosity may be examined in the case of ULXs whose spectra are well fitted by an absorbed disk black body  \citep[e.g.][]{Kajava2009, Kaaret2009}.  For some of the ULXs, the correlation between $\Gamma$ and luminosity  is positive and for some it is negative.  ULXs that become spectrally harder when they are in a brighter state have been found earlier \citep{Roberts2006,Feng2009,Gurpide2021}.  \citet{Ghosh2022a} also have found an anti-correlation between $\Gamma$ and luminosity for NGC 1042 ULX--1 (a stellar mass super-Eddington accretor).  
\citet{Kajava2009} have found four ULXs whose spectra can be fitted by a MCD model, showing a positive correlation between luminosity and disk temperature (same as Galactic Black Hole Binaries), with the luminosity-temperature diagram of the soft excess from the powerlaw plus MCD model fits following roughly a 
$L_{\rm soft} \propto T_{in}^{-3.5}$ relation.   \citet{Gurpide2021} have investigated, the correlation between the bolometric luminosity of the cool disc component and its temperature, for five ULXs from different galaxies, and observed a positive  correlation  $L_{\rm soft} \propto T_{in}^{\alpha}$; $\alpha$ values range from 1.27 to 3.55.  However, \citet{Miller2013} has obtained results which support a $L \propto T^4$ behaviour which indicates a Standard Shakura-Sunyaev Disk (assuming a constant inner radius) which radiates efficiently and the accretion is sub-Eddington \citep{Shakura1973}.  \citet{Watarai2001} introduced a slim disk model which provides a more self-consistent explanation for the spectral and temporal properties of ULXs in the super-Eddington accretion regime.  When the accretion rate exceeds a critical value, advection becomes significant, and energy is transported inward rather than radiated away locally. In the slim disk, the effective inner radius can appear larger or change with accretion rate, which affects how L varies with T \citep{Straub2013}.  Super-Eddington accretion can drive massive, optically thick winds from the disk. These winds can either obscure the inner disk or truncate the disk resulting in flattening or breaking of the L - T relation \citep{Koliopanos2017, Barra2022}.  NS-ULXs, exhibit distinct behaviors compared to BH ULXs due to their solid surfaces, strong magnetic fields \citep[e.g.][]{Israel2017a}, and boundary layer dynamics \citep[e.g.][]{Bachetti2014}. These factors alter the accretion geometry, radiative efficiency, and spectral properties, leading to deviations in the disk temperature–luminosity relation.\par 
\begin{figure*}
   \centering
    \includegraphics[scale=0.5]{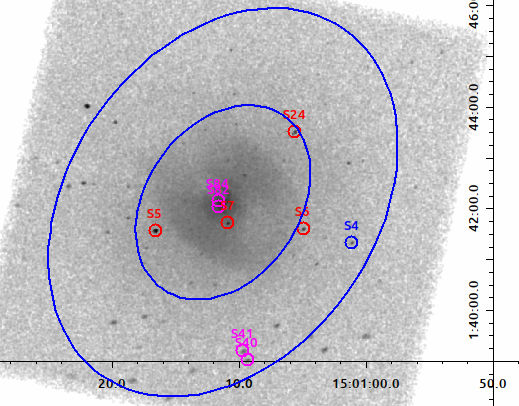}
    \caption{\textit{Chandra} ACIS X-ray image of NGC~5813 using the longest exposure observation (Obs Id:12952).  $D_{25}$ and $2\times D_{25}$ isophotal ellipses and candidate ULXs marked.  Red circles: Persistent ULX candidates identified by \citep{Liu2011} and re-identified here.  Blue circle: persistent ULX candidate newly identified in this study.  Magenta circles: ULX candidates identified by \citep{Liu2011} which turned out to be foreground sources or transient ULX candidates. }
    \label{fig:my_label_fig1}
\end{figure*}

\subsection{The ULXs in NGC~5813}
NGC~5813 is a central dominant (cD) giant elliptical galaxy with regular morphology, $31.67\pm 0.76$~Mpc away \citep{Tully2013} from us.   It is a member of the NGC 5846 group in the Virgo supercluster.  It shows evidence for a multiphase interstellar medium (ISM) with significant emission in cooling lines, H alpha, and X rays \citep{Werner2014}.  The galaxy harbours a kinematically distinct core (KDC) \citep{Emsellem2004, Krajnovi2015}, and  hosts both red and blue GC populations \citep{Hempel2007}.  \par
In the \textit{Chandra} ACIS observations of the galaxy, within the diffuse X-ray emission, three pairs of nearly collinear cavities at 1~kpc, 8~kpc, and 20~kpc from the central source are visible \citep{Randall2011, Randall2015}.  Two population studies of X-ray point sources have reported ULXs in NGC~5813. \citet{Liu2011}, in a \textit{Chandra} ACIS survey of X-ray point sources in 383 nearby galaxies,  identified 8 ULXs in NGC~5813; five ULXs within the D25 isophote and three in the area between the D25 and 2D25 isophotes (with the threshold luminosity for ULXs taken as $2.0 \times 10^{39}$ erg/s (0.3 - 8.0 keV)). The five ULXs within the D25 isophote were re-identified by \citet{Wang2016}; those with luminosity greater than or equal to $2.0 \times 10^{39}$ erg/s (0.3--8.0 keV) were considered.  These detections were based on a single \textit{Chandra} observation with obs id 5907 (see table \ref{tab:my_table_2}).  With more deep \textit{Chandra} observations as well as XMM-Newton observations of NGC~5813 at multiple epochs now available, there is good scope for a more detailed spectral and variability study of the ULXs of NGC~5813.\par

 \begin{table*}
     	
	\begin{tabular}{lcccr} 
		\hline
		  Source  & RA & DEC & Distance from  & Remark \\
		& h:m:s & d:m:s & the centre (arcmin) &  \\
		\hline
    
    CXOJ150110.877+014142.65 (S7) & 15:01:10.8814 & +1:41:42.823 & 0.430 &  re-identified$^*$ \\
    CXOJ150116.555+014133.97 (S5) & 15:01:16.5601 & +1:41:34.072 & 1.453 & re-identified$^*$ \\
	CXOJ150104.927+014136.02 (S6) & 15:01:04.9112 & +1:41:36.138 & 1.656 & re-identified$^*$ \\
	CXOJ150105.592+014330.76 (S24) & 15:01:05.5885 & +1:43:30.824 & 1.967 & re-identified$^*$ \\
    CXOJ150101.11+014119.80 (S4) & 15:01:01.1100 & +1:41:19.802 & 2.649 & newly identified \\
    
	   \hline
	\end{tabular}
    \caption{The candidate persistent ULXs in NGC~5813. The positions of the sources are from the longest observation Obs Id: 12952.  $^*$ Identified by  \citet{Liu2011} -- the source names used for these are the same as those used by \citet{Liu2011}. The sources are listed in the order of increasing radial distance from the centre.}
    \label{tab:my_table_1}
 \end{table*}
 
\section{Data Preparation and Analysis} 
\textit{Chandra} observations of NGC~5813 were used to identify the ULXs.  Data, from nine \textit{Chandra} Advanced CCD Imaging Spectrometer (ACIS) pointed observations of NGC~5813, were analyzed to identify the ULXs in the galaxy (details of the data are given in Table~\ref{tab:my_table_2}).  \textit{Chandra} has very good spatial resolution in the  0.3--10.0 keV energy band, which makes \textit{Chandra} the best instrument to study ULXs in NGC~5813 (because of its distance).  CIAO 4.10 was used for data processing.  All the \textit{Chandra} observations were reprocessed using \textit{Chandra repro} script. Background light curves were produced by the \textit{dmextract} tool, and background flares were removed using the thread available at \url{http://cxc.harvard.edu/ciao/threads/acisbackground}.  Point sources were identified by the \textit{wavdetect} tool with scales from 1 to 16 in steps of $\sqrt{2}$, detection threshold \textit{sigthresh} = $10^{-6}$,  on the cleaned event file of each of the observations.  The flux in the  0.3 -- 8.0 keV band, from each of these sources, was estimated by fitting an absorbed powerlaw model using the CIAO \textit{srcflux} tool.\par
\subsection{Sample Selection}  
Altogether, 132 X-ray point sources, from the nine \textit{Chandra} observations, have been detected within the 2D25 ellipse of NGC~5813.  For the distance of NGC~5813, a flux of $8.40\times 10^{-15}$ erg/cm$^2$/s corresponds to $10^{39}$ erg/s.   Sources, detected in all nine \textit{Chandra} observations, and having average luminosity greater than or equal to $10^{39}$ erg/s, were identified first.  There are eleven such sources after removing the known foreground and background sources.  The sample for the present study was then identified as follows. The spectra of these eleven sources were extracted from the nine \textit{Chandra} observations, in order to identify the persistent ULXs.  Five sources, all of which have luminosity above $10^{39}$ erg/s in all the nine \textit{Chandra} observations for all three models used, were identified and constitute the sample for the present study (See \S2.2 for details).    There were a few sources with luminosity above $10^{39}$ erg/s for at least one of the nine observations, but not all.  Such candidate transient ULXs will be studied in a later paper. \par 

 \begin{table}
	\centering
	
	\begin{tabular}{lllc} 
		\hline
		Obs Id & Detector & Date & Cleaned Exposure \\ & & & Time (ks)\\
		\hline
		5907 & ACIS-S & 2005 April 2 & 48.4\\
		9517 & ACIS-S & 2008 Jun 5 & 98.8\\
		12951 & ACIS-S & 2011 March 28 & 72.9\\
		12952 & ACIS-S & 2011 April 5 & 142.8 \\
		12953 & ACIS-S & 2011 April 7 & 31.7  \\
		13246 & ACIS-S & 2011 March 30 & 45.0 \\
		13247 & ACIS-S & 2011 March 31 & 35.5 \\
		13253 & ACIS-S & 2011 April 8 & 118.0 \\
		13255 & ACIS-S & 2011 April 10 & 43.4 \\
		\hline
	\end{tabular}
    \caption{Summary of \textit{Chandra} ACIS pointed observations of NGC~5813 }
	\label{tab:my_table_2}
 \end{table}
 
 \subsection{\textit{Chandra} Data: Spectral Analysis}
Cleaned event files of all the \textit{Chandra} observations listed in Table~\ref{tab:my_table_2} were used to extract spectra.  The diffuse emission present in the galaxy is not spatially uniform; also, there are three pairs of cavities in them. It is necessary therefore to characterize the diffuse emission around each source first, as the emission from the ULX arrives combined with this diffuse emission. The  extraction of the spectra is done using the CIAO tool \textit{specextract}, and modelling and fitting is performed using XSPEC version 12.12.0.\par

\subsubsection{The Diffuse Emission} \label{The Diffuse Emission}
For each point source, the spectrum of the diffuse emission was extracted, using the flux from an annular region around it  with inner radius $\sqrt{3} \times$ r$_7$ and outer radius 2 $\times$ r$_7$, where r$_7$ is the radius of a circle that includes 90 percent of the source flux at 7.0 keV at the position of the source. The background spectrum is extracted from four circular regions (radius = $4 \times$ r$_7$) randomly selected outside the 2$\times$D$_{25}$ region of the galaxy but on the same CCD as the galaxy.  For the innermost source S5, the diffuse emission spectrum in each observation could be modelled on its own. For the other sources, the surrounding diffuse emission did not have sufficient counts for individual modelling, so the spectrum obtained by stacking the diffuse emission spectra from all the nine observations was used to model the diffuse emission.  We used the CIAO tool "combine-spectra", which sums multiple PHA spectra, for this. This task also combines the associated background PHA spectra and source and background ancillary (ARF) and response (RMF) files.\par 
The diffuse emission spectra were fitted in the 0.3 -- 3.0 keV band with the Tbabs*Apec model\footnote[1]{Apec is an emission spectrum model from collisionally-ionized diffuse gas (\url{https://heasarc.gsfc.nasa.gov/xanadu/xspec/manual/XSmodelApec.html}) and Tbabs is the Tuebingen-Boulder ISM absorption model (\url{https://heasarc.gsfc.nasa.gov/xanadu/xspec/manual/node268.html})}.  The value of the absorption column density (n$_H$)  is fixed to the Galactic value (n$_H$ = 4.37$\times$ 10$^{20}$ cm$^{\minus2}$; \citet{Kalberla2005}). Abundance and plasma temperature values are obtained from the abundance and plasma temperature maps from \citet{Randall2015} and redshift = 0.00653 \citep{Cappellari2011}.  These parameters were fixed and the normalisation parameter of Apec was allowed to vary. The best-fit value of the Apec normalisation for the diffuse emission around each source is given in Table~\ref{tab:my_table_3}.\par

 \begin{figure}
    \centering
    \includegraphics[scale=0.5]{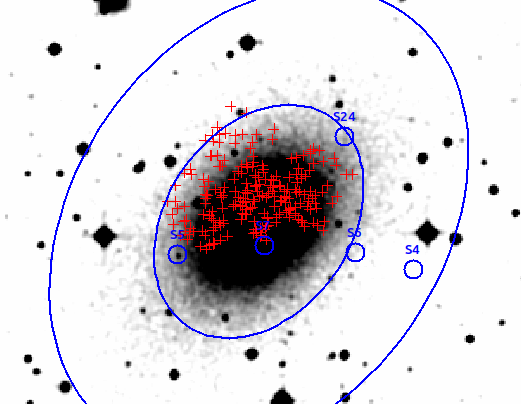}
    \caption{DSS image of NGC~5813. ULX candidates are showcased as blue circles, globular clusters \citep{Kundu2001} as red crosses }
    \label{fig:my_label_fig2}
\end{figure}
\begin{figure}
    \centering
    \includegraphics[scale=0.5]{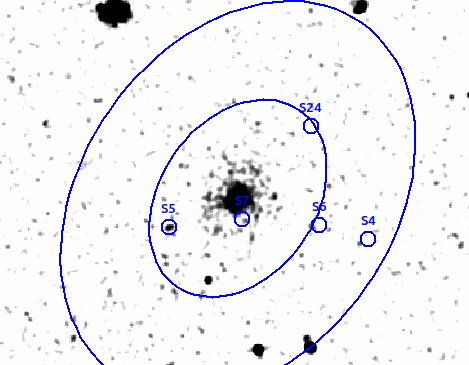}
    \caption{GALEX FUV observation of NGC~5813 with ULX candidates showcased as blue circles}
    \label{fig:my_label_fuv}
\end{figure}
 \begin{table*}
	\centering   
  
	\begin{tabular}{lcccr} 
		\hline
        Source &  Abundance & Plasma Temperature & Normalisation \\
        & &  (kT in keV) & ($\times$ 10 $^{-6}$) \\
        & (Literature$^a$) & (Literature$^a$) & (Estimated) \\
        \hline
        CXOJ150110.877+014142.65 (S7) & 0.36 & 0.57 & $^b$7.41 $_{-1.34}^{+1.34}$ \\
        \\
        CXOJ150116.555+014133.97 (S5)& 0.5 & 0.66 & 1.16$_{-0.16}^{+0.16}$\\
        \\
        CXOJ150104.927+014136.02 (S6) & 0.5 & 0.76 & 0.66$_{-0.11}^{+0.11}$ \\ 
        \\
        CXOJ150105.592+014330.76 (S24) & 0.58 & 0.76 & 0.33$_{-0.06}^{+0.07}$ \\
        \\
		CXOJ150101.10+014119.89 (S4) & 0.47 & 0.76 & 0.38$_{-0.09}^{+0.09}$ \\
        \\
		
		\hline
	\end{tabular}
    \caption {Parameters for the diffuse emission around the sample sources: the abundance is given in solar units. Unit of Apec normalisation is $10^{-14}cm^{-5}$.  $^a$ \citep{Randall2015}. $^b$ For this source alone, the value quoted is the average over nine individual observations and the error quoted is the standard deviation}. 
    \label{tab:my_table_3}
 \end{table*}

\subsubsection{The ULXs}
Energy spectra of the candidate ULXs were extracted from all the observations.  Source regions were demarcated by circles with radius r$_7$ in general.  For sources which are very close to each other, circles of radius r$_2$ (radius of a circle that includes 90 percent of the source flux at 2.0 keV at the position of the source) were used to avoid overlapping of the source regions.  As mentioned above, background spectra were extracted from circular regions (radius = $4 \times$ r$_7$) randomly selected outside the 2$\times$D$_{25}$ region of the galaxy but on the same CCD as the galaxy.  The spectra of the candidate ULXs were fitted with models such as:\\
\\ Model 1 : Tbabs $\times$ (Apec+powerlaw) \\
 Model 2 : Tbabs $\times$ (Apec+bbody) \\ Model 3 : Tbabs $\times$ (Apec+diskbb).\\\\ 
The parameters of the Apec model are frozen at the values given in Table~\ref{tab:my_table_3}. The absorption column density n$_H$ is frozen at Galactic absorption as mentioned in \S2.2.1.  After this,  more complex models, like :\\
\\
Tbabs $\times$ (Apec+bbody+powerlaw) \\
Tbabs $\times$ (Apec+diskbb+powerlaw) \\
Tbabs $\times$ (Apec+bbody+diskbb) \\
Tbabs $\times$ (Apec+diskbb+diskbb) \\ were fitted to the spectra of the ULXs.\par  

 \begin{table*}
    \centering
    \begin{tabular}{l|l|l|l|l|l|l|l|l|l|l}
    \hline
     
    \multirow{2}{*}{Source} & \multirow{1}{*}{Obs Id} &
      \multicolumn{3}{c|}{$\chi ^2/dof$} &
      \multicolumn{3}{c|}{L (10$^{+39}$erg/s)} &
      \multicolumn{3}{c}{Parameter} \\
       & & Model 1 & Model 2 & Model 3 & Model 1 & Model 2 & Model 3 & $\Gamma$ & kT (keV) & T$_{in}$ (keV) \\
        \hline
        S7 & 5907 & \textbf{4.65/7}  & 9.76/7  & \textbf{5.66/7} &  \textbf{3.02$^{+0.72}_{-0.67}$} &  2.13$_{-0.60}^{+0.54}$ & \textbf{2.62$^{+0.83}_{-0.69}$} & 1.30$^{+0.41}_{-0.39}$ & 0.86$^{+0.21}_{-0.17}$ & 1.99$^{+2.43}_{-0.71}$\\
        S7 & 9517 & \textbf{17.63/12} & 37.84/12 & 25.32/12 & \textbf{2.91$^{+0.40}_{-0.40}$} & - & - & 1.80$^{+0.25}_{-0.24}$ & - & -  \\
        S7 & 12951 & \textbf{6.30/8} & 12.63/8 & 8.48/8 & \textbf{3.71$^{+0.56}_{-0.56}$} & 1.80$^{+0.28}_{-0.28}$ & 2.26$^{+0.35}_{-0.34}$ & 2.16$^{+0.39}_{-0.39}$ &   0.37$^{+0.07}_{-0.06}$ &  0.59$^{+0.20}_{-0.13}$   \\
        S7 & 12952 & \textbf{23.80/21} & 86.91/21 & 52.38/21 & \textbf{4.16$^{+0.42}_{-0.42}$} & - & -  & 2.07$^{+0.18}_{-0.17}$ & - & - \\
        S7 & 12953 & \textbf{6.76/7} & 21.58/7 & 14.00/7 & \textbf{5.69$^{+1.04}_{-0.96}$} & - & 3.45$^{+0.61}_{-0.59}$ & 2.30$^{+0.40}_{-0.38}$ & - & 0.64$^{+0.25}_{-0.19}$  \\
            \hline

    \end{tabular}
    \caption{Model comparison of the sample sources using \textit{Chandra} observations.  $\chi ^2/dof$ and luminosity for the best fit model/s is given in bold.  If, for a model the reduced $\chi^2$ is greater than two, the luminosity and model parameters are not given here. The full version of this table is available online.}   
    \label{tab:Extra_tab1}
\end{table*}
\subsubsection{Luminosity Calculation}
The unabsorbed flux in the 0.3–8.0 keV band was estimated for each ULX using the cflux model. Table~\ref{tab:Extra_tab1} gives the parameters of Model 1, Model 2, and Model 3, fitted to the spectra of each of the sources for the nine \textit{Chandra} observations, along with the luminosity of each source.  Sources S4, S5, S6, S7, and S24 have luminosity above 10$^{39}$ erg/s in all nine \textit{Chandra} observations for all three models.  Hence, they are classified as persistent ULXs.  We set this criterion because unabsorbed luminosities from powerlaw model are often overestimated since at lower energies, powerlaw does not have a cut-off.  \citet{Liu2011} identified eight candidate ULXs with luminosity above 2$\times$10$^{39}$ erg/s (0.3–8.0 keV) in the 2005 \textit{Chandra} ACIS observation of NGC~5813; four of them, S5, S6, S7 and S24, are re-identified as persistent ULXs in this study.  Two sources, CXOJ150111.491+014202.74 (S82) and CXOJ150111.645+014207.42 (S84), are not detected consistently  in all the nine \textit{Chandra} observations.  The remaining two, CXOJ150109.329+013900.72 (S40) and CXOJ150109.660+013911.63 (S41), were identified as eclipsing binaries \citep{Khrutskaya2004} and are foreground sources.  S4, is here newly identified as a persistent ULX (Figure ~\ref{fig:my_label_fig1}).

\begin{table}
    \centering
    \begin{tabular}{cccc}
    \hline
         Obs Id & Instrument and & Date & Live time for  \\
         & Filter & & the central CCD (for EPN) \\
         \hline
         0302460101 & EPIC, Medium & 23/07/2005 & 21.06 ks  \\
         0554680201 & EPIC, Thin & 11/02/2009 & 38.75 ks  \\
         0554680301 & EPIC, Thin & 17/02/2009 & 35.28 ks  \\
         \hline
    \end{tabular}
    \caption{XMM-Newton observations of NGC~5813 used in this paper}
    \label{tab:my_table_4}
    
\end{table}

\subsection{XMM-Newton Data: Analysis}
Archival data from three XMM-EPIC observations of NGC~5813 were included in our study (details are given in Table~\ref{tab:my_table_4}), to obtain the state of the candidate ULXs during three additional time epochs. Data analysis was carried out using the software SAS version 21.0.0. In each case, the observation data file (ODF) was downloaded and the data reprocessed using \textit{epproc} and \textit{emproc} commands for EPIC-pn and EPIC-mos instruments, respectively.  The EPIC event list was filtered for periods of high background flaring activity \footnote[3]{\url{(https://www.cosmos.esa.int/web/xmm-newton/sas-thread-epic-filterbackground-in-python)}}.  A Good Time Interval (GTI) file, where the background count rate is low and steady above 10.0 keV for EMOS, and between 10.0 and 12.0 keV for EPN, was then created.  The live time for the central CCD is obtained from the cleaned event file. The SAS source detection script \textit{edetect chain} was used to simultaneously search for sources in 5 images extracted in the 0.2-0.5 keV, 0.5-1 keV, 1-2 keV, 2-4.5 keV, 4.5-12 keV energy bands. This detection chain was separately run for each of the 3 EPIC cameras\footnote[4]{\url{(https://www.cosmos.esa.int/web/xmm-newton/sas-thread-src-find)}}.\par

We could not detect the source S7 in the XMM-Newton observations 0302460101 and 0554680201.  This is probably because S7 is just $0.43^\prime$ away from the centre of the galaxy, that is, inside the point spread function (PSF) of the central source of NGC~5813.  We could detect S7 in the 0554680301 MOS2 observation, but none of the three spectral models gave a good fit, probably because of heavy contamination from the central source.  The source S5 is detected in all observations.  S6 is detected in all observations except 0554680201 EPN and 0554680201 EPN where S6 falls in a chip gap.  S24 is detected in all observations except 0302460101 EPN, where it falls in a chip gap.  Also, in \textit{Chandra} observations of S24, we can see a nearby bright X-ray point source, which is resolved in \textit{Chandra} observations but cannot be resolved in XMM-Newton observations.  So, we have not included the spectral analysis of XMM-Newton observations of S24 either (Fig.~\ref{fig:my_label_lc}) in this study.  We could detect S4 in the 0554680301 MOS2 observation only.  There is a possibility that this source may be a transient.  But, since the PSF of XMM-Newton is comparatively larger and the exposure times of these XMM-Newton observations are short, it is also possible that this source may just not be bright enough to be detected in these observations.  In fact, of the remaining six out of the eleven sources selected at stage one of the selection process (\S2.1), five are not detected in any of the XMM-Newton observations.  The spectral analysis of sources S5, S6 and S4 were done for the observations in which these sources were detected. 
\subsubsection{Spectral Analysis: The Diffuse Emission}
To study the diffuse emission, we extracted spectra of the emission around the ULXs using annular regions of inner radius $\sqrt{3}$ times and outer radius two times the radius of the region from which the ULX spectrum is extracted: 25.98$^{\prime\prime}$ and 33$^{\prime\prime}$ respectively.  When there is a source adjacent to the ULX, a circular region close to the ULX region (the cavities present were avoided), of the same radius as the ULX region and also at the same distance from the centre of the galaxy was used to study the diffuse emission.   The source and background region areas used to make the spectral files were calculated using \textit{backscale}\footnote[5]{\url{(https://www.cosmos.esa.int/web/xmm-newton/sas-thread-mos-spectrum)}}.  \textit{Rmfgen} and \textit{arfgen} were run to create the redistribution matrix and the ancillary file, respectively.  To link the spectrum and associated files and to rebin the spectra, the task \textit{specgroup} was used.  The spectrum was rebinned such that there are at least 25 counts for each spectral channel\footnote[6]{\url{( https://www.cosmos.esa.int/web/xmm-newton/sas-thread-mos-spectrum)}}. 
The spectral analysis of diffuse emission is performed as described in \S2.2.1, except that we model the spectrum diffuse emission around each ULX separately for each observation and each instrument.  Also the plasma temperature parameter and the Apec normalisation for the diffuse emission are allowed to vary, but abundance is fixed to the area averaged value obtained from abundance map of NGC~5813 from \citet{Randall2015}. 
\subsubsection{Spectral Analysis: The ULXs}
The spectra of the candidate ULXs were extracted from circular regions of 16.5$^{\prime\prime}$ radius. The background spectra were obtained from circular regions of 32$^{\prime\prime}$ radius in a source-free region on the same CCD as the source region. 
The ULX spectra were fitted with three different models as outlined in Table ~\ref{tab:Extra_tab2}, with parameters nH, and redshift values the same as mentioned in \S\ref{The Diffuse Emission}.  Since the source regions and the diffuse emission regions are larger in size for XMM-Newton observations than for \textit{Chandra} observations, the plasma temperature and abundance parameter values have wider variation across the selected regions, and area averaged values, determined using the plasma temperature and abundance maps of NGC~5813 \citep{Randall2015}, were taken. For the diffuse component, the Apec normalisation is fixed to the value obtained from the diffuse emission fit described above.\par

In the case of S6, it is seen that the value of the photon index obtained from the powerlaw fit in XMM-Newton observations is higher than the values obtained from Chandra observations; the error bars overlap for some of the observations.  In the case of black body temperature, the values obtained are lower compared to Chandra observations; here also, not all error bars overlap.  For this source, none of the models tried, including complex models, gave a good fit for 0302460101 (PN) and 0554680301 (MOS) observations.  In the case of S5, the black body temperature values obtained are higher than those obtained from the analysis of Chandra data; the error bars mostly overlap.  The photon index values obtained are within errors the same.  For source S4 only one model yielded a good fit with XMM-Newton data. The value obtained for the black body temperature is lower compared to the results from Chandra observations and it is outside the error bars.  For details see Table ~\ref{tab:Extra_tab2}.\par

\begin{table*}
\centering

    \begin{tabular}{l|l|l|l|l|l|l|l|l|l|l|l}
    
    \hline
     
    \multirow{2}{*}{Source} & \multirow{1}{*}{Obs Id} & \multirow{1}{*}{Apec norm} &
      \multicolumn{3}{c|}{$\chi ^2/dof$} &
      \multicolumn{3}{c|}{L (10$^{+39}$erg/s)} &
      \multicolumn{3}{c}{Parameter} \\
     & & (10$^{-5}$) & Model 1 & Model 2 & Model 3 & Model 1 & Model 2 & Model 3 & $\Gamma$ & kT (keV) & T$_{in}$ (keV) \\
    \hline
    S5 & 101(MOS) & 4.53 & \textbf{31.68/36} & 56.03/36 & 42.00/36 & \textbf{12.32$^{+1.59}_{-1.57}$} & 6.71$^{+1.07}_{-1.01}$ & 7.82$^{+1.04}_{-1.02}$ & 2.09$^{+0.22}_{-0.24}$ & 0.28$^{+0.05}_{-0.04}$ & 0.50$^{+0.13}_{-0.09}$ \\
    S5 & 101(PN) & 3.41 & 65.32/26 & 86.74/26 & 79.58/26 & -& -& - & - & - & -\\
    S5 & 201(MOS) & 4.42 & \textbf{105.78/68} & 237.36/68 & 185.21/68 & \textbf{12.37$^{+1.11}_{-1.11}$} & - & - & 2.15$^{+0.12}_{-0.13}$ & - & - \\
    S5 & 201(PN) & 4.25 & \textbf{56.03/40} & 165.54/40 & 124.73/40 & \textbf{12.99$^{+1.10}_{-1.09}$} & - & - & 2.14$^{+0.10}_{-0.09}$ & - & - \\ 
    S5 & 301(MOS) & 3.65  & 141.60/65 & 178.17/65 & 168.74/65 & -& -& - & - & - & -\\
        
    \hline
    \end{tabular}
    
    \caption{Spectral analysis of ULXs: XMM-Newton observations.  Obs Id 101 : Obs Id 302460101, Obs Id 201 : Obs Id 554680201, and Obs Id 301 : Obs Id 554680301. $\chi ^2/dof$ and luminosity for the best fit model is given in bold. Source S7 and S4 are detected only in 554680301(MOS 2).  Unit of Apec normalisation is $10^{-14}cm^{-5}$. A full version of this table is available online.}
    \label{tab:Extra_tab2}
\end{table*}
\subsection{X-ray Colour-Colour Diagram}
X-ray colours provide information about the spectral state of the sources.  To calculate colors, photon counts were obtained for each candidate ULX for three energy bands, 0.5 keV -- 1.7 keV (Soft), 1.7 keV -- 2.8 keV (Medium), and 2.8 keV -- 8.0 keV (Hard),  denoted as S, M and H respectively \citep{Albacete2007}.  X-ray colours, C$_1$ = (M-S)/(M+S) and C$_2$ = (H-M)/(H+M), and the errors on the colours were estimated \citep{Park2006}. 
For each source, for each observation (and for each detector separately for XMM-Newton observations), the diffuse emission and the background emission were determined (for details please see \S2.2.1 for \textit{Chandra} observations, and \S2.3.1 for XMM-Newton observations), and subtracted from the emission from the source region.
The sources S4, S5 and S6 had zero photon counts in M and H bands for observations 0554680301 MOS2, 0302460101 MOS1 and 0554680301 MOS2 respectively.  Also, The source S5 had zero counts in H band for 0302460101 MOS2 and source S6 had zero counts in H band for 0302460101 MOS1, 0302460101 MOS2 and 0554680201 MOS2.  So, these observations were not included in this analysis.  The Colour-Colour diagram is plotted with C$_1$ on Y-axis and C$_2$ on X-axis (Figure~\ref{fig:my_label_cc}).\par
In the Colour-Colour diagram, the relative positions of the sources are seen to follow a trend.  Mean value of C$_1$  is comparatively high, for S6 (-0.08), and S4 (-0.18), which are best fitted with Tbabs*(Apec+Bbody) model; their state is determined as 'broadened disc'.  Mean value of C$_1$ is comparatively low, for S5 (-0.61), and S7 (-0.64), for which the best fit model includes a powerlaw component; their state is determined as 'hard-ultraluminous'.  For details regarding the determination of the states of the sources please see \S2.9.  In the case of S6, the value of C$_1$ is low for 2009 XMM-Newton observation with obsid 0554680301(MOS2), for which none of the three basic models gives a best fit even though the total count is not low.  Again, for S6, the value of C$_1$ obtained is low for 2011 \textit{Chandra} observation with obsid 13247; the best fit model includes a powerlaw for this observation alone.  It was not possible to fit Tbabs*(Apec+Diskbb+Powerlaw) model to examine whether the state of the source had also suffered a change or not, since the number of counts is not enough. For source S24, the best fit model is Tbabs*(Apec+Bbody), and the mean value of C$_1$ is -0.35; the state of the source has been determined as 'broadened disc'.  

\subsection{Intra-observational Variability}
To assess the intra-observational variability of the candidate ULXs,  the Gregory-Loredo algorithm \citep{Gregory1992} implemented as the CIAO tool \textit{glvary} was used.  It splits the events into multiple time bins and looks for significant deviations between bins.  The tool assigns a variability index based on the total odds ratio, the corresponding probability of a variable signal, and the fractions of the lightcurve within 3 $\sigma$ and 5 $\sigma$ of the average count rate.  The Gregory-Loredo variability algorithm provides the probability that the calculated flux from the source region is not constant throughout the observation.  The algorithm determines variability based on an odds ratio for the inter-arrival times of the photons within the source region being not uniformly distributed in time for each energy band. The appropriate sampling distribution for data from X-ray detectors, some or all of which may be attributed to background events, is the Poisson distribution \citep{Gregory1992}.  The variability index obtained was less than or equal to two, for all the candidate ULXs, in all the \textit{Chandra} observations as well as in the XMM-Newton observations.  Hence, none of the candidate ULXs may be considered as variable within the time period  of each of the observations as per the interpretation of the Gregory-Loredo variability index.  A source is considered to be variable if the variability index is five or above.

\subsection{Inter-observational Variability}
\begin{figure*}
     \centering
     \includegraphics[width=0.6\columnwidth]{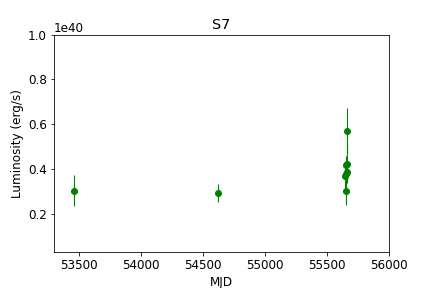}
     \includegraphics[width=0.6\columnwidth]{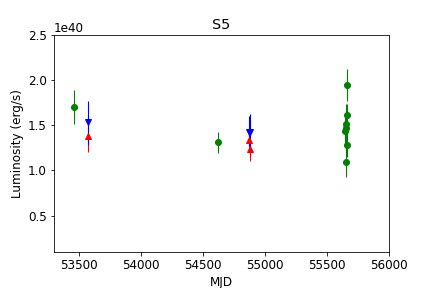}
     \includegraphics[width=0.6\columnwidth]{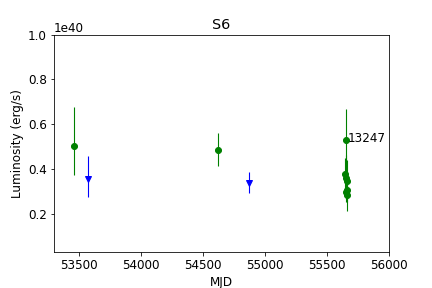}
     \includegraphics[width=0.6\columnwidth]{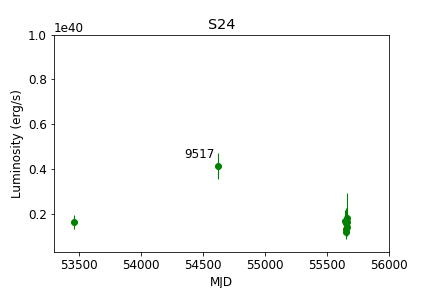}
     \includegraphics[width=0.6\columnwidth]{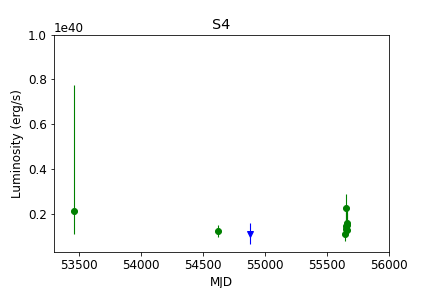}
     \includegraphics[width=0.2\columnwidth]{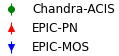}
     \caption{Lightcurves of the ULXs.  The best fit models given in Table ~\ref{tab:my_table_5} are used to estimate the luminosity.  For the ULX S24, the best fit model (Model 2) does not give good fit for the observation 9517 so, the model that gives the best fit for that observation i.e., Model 1 is used to estimate the luminosity for that particular observation. Similarly, for S6, the best fit model (Model 2) does not give good fit for the observation 13247 so, Model 1 which gives best fit for that observation is used to estimate luminosity for that particular observation.}
     \label{fig:my_label_lc}
 \end{figure*}

The lightcurves of the candidate ULXs, extracted from \textit{Chandra} and XMM-Newton observations, are given in Figure ~\ref{fig:my_label_lc}.  To explore the inter-observational variability of the ULXs, the unabsorbed flux (0.3 - 8.0 keV) from the spectrum of each source for each of \textit{Chandra} and XMM-Newton observations was used.  The fractional variability was calculated using the formula : \par
\begin{equation*}
    F_{var} = \frac{\sqrt{S^2 - \bar{\sigma}^2_{err}}}{\bar{x}}
\end{equation*}
where S$^2$ is the variance of flux values, $\bar{\sigma}$$^2_{err}$ is the mean square of errors on flux values, and $\bar{x}$ is the mean of the flux values \citep{Vaughan2003}. The error on the fractional variability was also estimated using the formula given by \citet{Edelson2002}.  The errors on the estimated $F_{var}$ are at 68\%.  We first obtained the fractional variability of the sources, between the seven observations made from 28 March 2011 to 10 April 2011 (seven \textit{Chandra} observations).  For sources S4, S6, and S24, the variance of flux values was smaller than the mean square of errors on the flux values and thus the fractional variability over the time scale of days for these sources could not be ascertained.  For S5, the fractional variability over the time scale of days is estimated to be 15.1$\pm 5.8\%$ and for S7 it is 10.8$\pm10.0\%$ which is compatible with absence of variability. \par
Then we estimated the fractional variability over the time scale of years.  We have \textit{Chandra} observations from 2005, 2008 and 2011 and XMM-Newton observations from 2005 and 2009.  There are multiple observations available from 2005, 2009 and 2011  and, for these years,  only the data from the longest observation was used for the analysis.  For S7, the estimated value of F$_{var}$ is 14.4$\pm12.0$\%, for S6 it is 12.5$\pm18.0$\%, for S24 it is 62.0$\pm27.0$\%, and for S5 it is 9.4$\pm5.8$\%.  We could not estimate the value for S4 because the variance of flux values was smaller than the mean square of errors on the flux values.

\subsection{Luminosity - \texorpdfstring{$\Gamma$}{Gamma} Correlation}
\citet{Kajava2009} and \citet{Kaaret2009} investigated the relationship between luminosity and photon index ($\Gamma$), for ULXs whose spectra are well fitted with an absorbed powerlaw model.   ULXs that hardened at high luminosities and those that softened at high luminosities have both been discovered.  The relation between $\Gamma$ and luminosity can give insights into the nature of the source.  Here, photon indices and luminosities of the ULXs, at various time epochs are determined.  All of the ULXs are well fitted with Model 1 except S24 for the 5907 \textit{Chandra} observation and S4 for the 0554680301 XMM-Newton EPIC-MOS observation.  For S5, we used Tbabs*(Apec+Bbody+Powerlaw) model to get the luminosity and photon index.  Kendall's $\tau$ rank correlation coefficient\footnote[7]{\url{(https://www.gigacalculator.com/calculators/correlation-coefficient-calculator.php)}} was  estimated for each source (Table ~\ref{tab:my_table_5}).  Except S4, no other sources show any significant correlation.  For S4, Kendall's $\tau$ correlation coefficient is 0.50 (p-value = 0.03).  This suggests that ULX S4 hardened at high luminosities.  For other sources the correlation is not statistically significant.  The luminosity and $\Gamma$ values have rather significant error bars. Deeper observations and more detailed analyses are needed in order to draw significant inferences. 
\subsection{%
  \texorpdfstring{$L_{\text{Soft}}$}{L\_Soft} - %
  \texorpdfstring{$kT_{\text{in}}$}{kT\_in} Correlation%
}
For the ULXs S4, S6, S7 and S24, Model 3 was fitted to the data to explore the correlation between the luminosity from the disk (L$_{Soft}$) and the inner disk temperature (kT$_{in}$).  Here, L$_{Soft}$ is obtained as the unabsorbed luminosity estimated from the Diskbb component for the energy range 0.3 - 8.0 keV.  Since S5 cannot be fitted well with Model 3, parameters obtained by fitting the Tbabs*(Apec+Diskbb+Powerlaw) model to its spectra was used.  Other than in the case of S24, no significant correlation is noted for these ULXs.  For S24, Kendall's $\tau$ correlation coefficient is 0.54 (p-value = 0.02).  However, since, the error bars for kT$_{in}$ and L$_{Soft}$ are large for most of the observations, no conclusions may be drawn regarding the correlation between inner disc temperature and luminosity of S24 either.
\subsection{The State of the ULXs}   
Usually, ULXs are observed in three accretion regimes: the Broadened Disc (BD) state, Hard Ultraluminous (HUL) state, and Soft Ultraluminous (SUL) state \citep{Sutton2013}.  To ascertain the state of the ULXs here, their spectra were fitted with a Tbabs*(Apec+Diskbb+Powerlaw) model and the values for the  inner disc temperature (kT$_{in}$), $\Gamma$ and the ratio of the unabsorbed flux contribution from the powerlaw component to the unabsorbed flux contribution from the diskbb component, for the energy range 0.3 to 1.0 keV ($F_{pl}/F_{disc}$), were obtained.  The different states are delineated by \citet{Sutton2013} as follows:\\
\\
kT$_{in}$ > 0.5 keV \& $F_{pl}/F_{disc}$ < 5  : BD state \\
kT$_{in}$ > 0.5 keV,  $F_{pl}/F_{disc}$ > 5, \& $\Gamma$ < 2.0 : HUL state \\
kT$_{in}$ > 0.5 keV,  $F_{pl}/F_{disc}$ > 5, \& $\Gamma$ > 2.0 : SUL state \\  
kT$_{in}$ < 0.5 keV \& $\Gamma$ < 2.0 : HUL state \\
kT$_{in}$ < 0.5 keV \& $\Gamma$ > 2.0 : SUL state \\

\begin{figure*}
    \centering
    \includegraphics[scale=0.32]{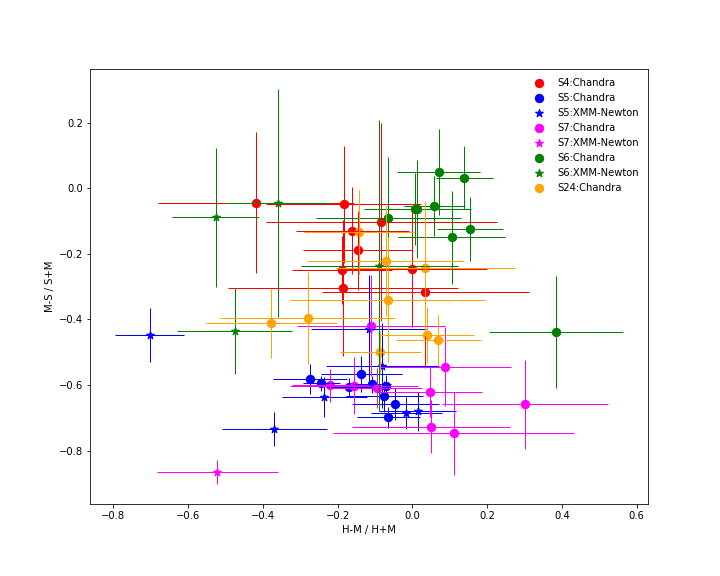}
    \caption{Colour--Colour diagram of the ULXs, from \textit{Chandra} observations and XMM-Newton observations. }
    \label{fig:my_label_cc}
\end{figure*}

The model fit was effected for the two longest \textit{Chandra} observations 12952 and 13253. Results are given in Table~\ref{tab:my_table_st}.  It is observed that the sources fall into two classes, those in the BD state and those in the HUL state.  Sources in these two different states are seen to  fall in separate regions in the Colour-Colour diagram  (Figure~\ref{fig:my_label_cc}).

\begin{table}
    \centering
    \begin{tabular}{c c c c c c c}
        \hline
         Source & Obs Id & kT$_{in}$ & $\Gamma$ & $F_{pl}/F_{disc}$ & State   \\
         & &  (keV) & & (0.3-1.0keV) & \\
         \hline
         S4 & 12952 & 1.33 & 1.75 & 0.01 & BD \\
         S4 & 13253 & no good fit & & & \\
         S5 & 12952 & 0.21 & 1.87 & - & HUL\\
         S5 & 13253 & 0.19  & 1.80 & - & HUL \\
         S6 & 12952 & 1.94 & 0.31 & 0.28 & BD \\
         S6 & 13253 & 2.36 & 0.32 & 0.40 & BD \\
         S7 & 12952 & 0.27 & 1.63 & - & HUL \\
         S7 & 13253 & 0.36 & 1.72 & - & HUL \\
         S24 & 12952 & no good fit & & & \\
         S24 & 13253 & 1.31 & 0.25 & 0.09 & BD \\
         \hline
    \end{tabular}
    \caption{Classification of the ULXs based on their state.}
    \label{tab:my_table_st}
\end{table}
\justify 

\section{Results}
\begin{table*}
    \centering
    \begin{tabular}{|c|c|c|c|c|c|c|}
    \hline
         Source & Best fit model$^a$ & Average L & Inter-observational  & L$_x$ - $\Gamma$ Kendall's $\tau$ $^b$ & L$_{Soft}$ - kT$_{in}$ Kendall's $\tau$ $^c$ & Counterpart  \\
         & & (10$^{39}$erg/s) & fractional variability & correlation coefficient & correlation coefficient & (FUV / Optical) \\
         \hline
         S7 & Tbabs*(Apec+powerlaw) & 3.82 & 10.8$\pm10.0\%$ & +0.28 (p = 0.15) & +0.14 (p = 0.32 )& Not detected \\
         
         S5 & Tbabs*(Apec+bbody+powerlaw) & 14.5 & 15.1$\pm5.8\%$ & +0.14 (p = 0.23) & +0.43 (p = 0.01) & FUV and Optical \\
         S6 & Tbabs*(Apec+bbody) & 3.80 & - & +0.16 (p = 0.24) & -0.02 (p = 0.46) & Not detected \\
         S24 & Tbabs*(Apec+bbody) & 1.83 & - & -0.26 (p = 0.18) & +0.54 (p = 0.02) & Not detected  \\
         S4 & Tbabs*(Apec+bbody) & 1.50 & - & -0.50 (p = 0.03) & +0.33 (p = 0.10) & Not detected  \\
         \hline
    \end{tabular}
    \caption{X-ray spectral, fractional variability, and UV/optical properties of ULXs. Average luminosity obtained using best fit model of the ULX. The inter-observational fractional variability is for the period from 28 March 2011 to 10 April 2011. $^a$ For majority of observations. $^b$ Model used: Tbabs*(Apec+powerlaw). $^c$ Model used: Tbabs*(Apec+Diskbb) for sources S7, S6, S24 and S4; Model used: Tbabs*(Apec+Diskbb+Powerlaw) for S5.}
    \label{tab:my_table_5}
\end{table*}
\subsection{CXOJ150110.877+014142.65 (S7)}
This ULX lies within the D25 ellipse of the galaxy and is the one closest to the centre of the galaxy among the five ULXs.  S7 is not visible in FUV.  Due to the diffuse emission in the galaxy, for this source, no conclusions can be drawn regarding any probable optical  counterpart, from the DSS image (Figure~\ref{fig:my_label_fig2}).  S7 is the only ULX for which a possible association with a globular cluster, in the catalogue prepared by \citet{Kundu2001},  may be surmised.  The nearest GC is 0.1' distance away from S7 and this GC has v-magnitude 23.86$\pm$0.093, i magnitude 22.86$\pm$0.17, and V-I Colour 1.0$\pm$0.19.  This is typical for GCs in elliptical and S0 galaxies and also indicates that the GC is having high metallicity \citep{Kundu2001b}.  The spectrum of the source is  quite soft.  Spectral analysis yields  Tbabs*(Apec + Powerlaw) as the preferred model for this source.  Average value of $\Gamma$ of the source is 1.89 $\pm$ 0.30 (Coefficient of variation (CoV) = standard deviation divided by mean is 15.9\%).    
There is no intra-observational or inter-observational variability for S7 (see section \S2.5 and \S2.6).  In the case of  XMM-Newton observations, due to its location within the PSF of the central source, this source is detected only in the  0554680301 EMOS2 observation.  Augmented models like bbody/diskbb added to Model 1 and Model 3 were also fitted to the spectra of this ULX but they are not statistically requested by the data.

\subsection{CXOJ150116.555+014133.97 (S5)}
S5 is one of the brightest X-ray sources in NGC~5813.  It lies inside the D25 ellipse.  The source has an optical (Figure~\ref{fig:my_label_fig2}) and a FUV counterpart (Figure~\ref{fig:my_label_fuv}).  The optical counterpart has u magnitude=20.52$\pm$0.08, g magnitude=20.90$\pm$0.04, r magnitude=20.62$\pm$0.05, i magnitude=19.51$\pm$0.03, z magnitude=19.19$\pm$0.07. \citep{Abazajian2009}.  In the FUV, Gaia magnitude is 20.94 $\pm$ 0.02 \citep{Gaia2020}.  We have estimated absolute magnitudes in the u, g, r, i, z filters of SDSS and estimated Colours: U-G Colour = -0.43$\pm0.12$, G-R Colour = +0.24$\pm0.10$, R-I Colour = +1.07 $\pm 0.09$ and I-Z Colour =  +0.30 $\pm0.11$.  Negative U-G suggests that the object is brighter in the  U band than in the G band, which can be  due to either X-ray reprocessing or accretion disk emission \citep{Tao2011} or the source could be a young early type OB star cluster \citep[e.g.][]{Liu2004} or it could be emission from the nebula surrounding the ULX \citep{Pakull2003}.  The red R-I Colour indicates either a Red Supergiant donor or line emission found in ULX nebulae.\par 
The extracted spectra of source S5 are well fitted by Model 1 only.  Model 2 and Model 3 give reduced $\chi^2$ greater than 2 (except for 0302460101 EMOS observation).  The average value of $\Gamma$ is 2.06 $\pm$ 0.08   (CoV =  3.9\%).  Model 1 and Model 3 augmented with bbody/diskbb were fitted to the spectra of this source.  We used  F-test and Akaike Information Criterion (AIC) to check if the augmented models significantly improve the fit. The value of AIC  for the Tbabs*(Apec+Bbody+Powerlaw) model is minimum for five out of fifteen (\textit{Chandra} and XMM-Newton) observations and $\Delta$AIC < 2 for eight observations and between 2  and 4 for just two observations. If $\Delta$AIC is  < 2, it suggests that the alternate model is nearly on par with the best model. If $\Delta$AIC  ranges between 4 and 7, the support for the new model is notably weaker. A $\Delta$AIC  > 10 implies the new model is unlikely to be the best model \citep{Akaike1974}.  So, we can say that the addition of bbody to Model 1 really improves the fit.  Fitting the spectra of S5 with Model 1 augmented with bbody yields kT in the range 0.10 - 0.20 keV and $\Gamma$ in the range 0.97 - 1.93.  The average apparent emitting radius of the bbody component is 6900 $\pm 1758$ km (excluding outliers).   Model Tbabs*(Apec+Diskbb+Diskbb) gave minimum AIC value for \textit{Chandra} observations 13246, 13253 and 13255 but for eight observations $\Delta$AIC has values ranging from 2 to 16.  So we can not say this model is superior to Tbabs*(Apec+Bbody+Powerlaw).  The $T_{in}$ value for soft diskbb varies from 0.13 to 0.32 and that for harder diskbb varies from 1.04 to 4.73.  
This source exhibits 15.1$\pm5.8\%$ inter-observational fractional variability over the time scale of days but there is no intra-observational variability (see \S2.5 and \S2.6).  
\subsection{CXOJ150104.927+014136.02 (S6)}
ULX S6 lies just outside the D25 ellipse of NGC~5813.  Neither optical nor FUV counterparts can be identified in DSS or GALEX FUV images for this ULX.  For the deepest \textit{Chandra} observation, Model 2 and Model 3 are giving good fit and Model 1 is giving a bad fit.  Also,  the spectrum of S6 is well fitted with Model 2, the absorbed blackbody, for more of the observations than it is with other models.  So, we are taking Model 2 as the best fit model for S6.  The average value of 
black body temperature is 0.86 $\pm$ 0.24 (CoV = 27.5\%).  The average apparent emitting radius of the bbody component is 190 $\pm 36$ km.  This ULX is comparatively harder than the other ULXs in NGC~5813 with $\Gamma$ values less than 1.0 in all nine \textit{Chandra} observations. Though in XMM-Newton observations, $\Gamma$ values are greater than 1.5  for most of the observations, the spectrum of S6 is well fitted with Model 2 which is the absorbed Bbody. The hardness of the spectra of this ULX may be an indication of the possibility of this ULX being a PULX \citep{Gurpide2021b}.  A detailed study of this source is required to confirm this.  Model 1 augmented with bbody/diskbb and Model 3 augmented with a bbody/diskbb do not improve the fit.  There is no intra-observational or inter-observational variability for S6 (see \S2.5 and \S2.6). 
\subsection{CXOJ150105.592+014330.76 (S24)}
S24 lies on the D25 isophotal ellipse.  Neither optical nor FUV counterparts could  be found in DSS or GALEX observations for this source.  In most of the observations Model 2 gives the best fit and for the deepest \textit{Chandra} observation, Model 2 is giving a good fit.  So, we consider Model 2 (absorbed Bbody) as the best-fit model for S24.  The average value of blackbody temperature is 0.72 $\pm$ 0.12 keV (CoV = 17.4\%).  The average apparent emitting radius of the black body component is 250 $\pm  94$ km.  Model 1 augmented with bbody/diskbb and Model 3 augmented with bbody/diskbb were also fitted to the spectra of S24.  The augmented models are not statistically requested by the data.  There is no intra-observational or inter-observatioal variability for this source (see \S2.5 and \S2.6).  There is possibly a moderate positive correlation between inner disk temperature (kT$_{in}$) and L$_Soft$ for this ULX but the error bars are very large and we can not confirm this correlation.
\subsection{CXOJ150101.10+014119.89 (S4)}
S4 lies outside the D25 ellipse but within the 2D25 ellipse.  It is not visible in DSS and GALEX observations.  Though the spectrum of the source can be fitted with Model 1, Model 2 and Model 3, for the deepest \textit{Chandra} observations Model 2 and Model 3 are giving a good fit.  Also, for majority of observations Model 2 gives the best fit.  So, we consider Model 2 (absorbed Bbody) as the best-fit model for S4.  The average value of black body temperature is 0.70 $\pm$ 0.17 keV (CoV = 24.5\%).  The average apparent emitting radius of the bbody component is 200$\pm  40$ km.  Model 1 augmented with bbody/diskbb and Model 3 augmented with a bbody/diskbb do not improve the $\chi^2$ statistic.  There is possibly a moderate negative correlation between luminosity and $\Gamma$ for this source; however the error bars are large and we cannot confirm this correlation.  There is no intra-observational or inter-observatioal variability for this source (see \S2.5 and \S2.6).

\section{Discussion}
From the analysis, five ULXs, each having luminosity consistently greater than 1.0 $\times$10$^{39}$ erg/s across the nine \textit{Chandra} observations for all the three models fitted have been identified and out of these five sources, CXOJ150116.555+014133.97 (S5), CXOJ150110.877+014142.65 (S7) and CXOJ150105.592+014330.76 (S24) are within the D25 ellipse while CXOJ150104.927+014136.02 (S6) and CXOJ150101.10+014119.89 (S4) are in the region between the D25 and 2D25 ellipses.  Out of the five ULXs in NGC~5813, S5 has average luminosity > 10$^{40}$ erg/s, more than three times the average luminosity of the next brigthest source S7 (average luminosity above 3.0 $\times$ 10$^{39}$  erg/s).  The source S6 has average luminosity above 3.0 $\times$ 10$^{39}$  erg/s.\par 
None of the ULXs in NGC~5813 show intra-observational variability. Sources S7, S24, S6 and S4 are not significantly variable over the observations separated by days, with  only S5 showing an inter observational fractional variability of 15.1$\pm5.8\%$.  There is no unambiguous evidence of longer term (over years) variability for these ULXs.  Generally, ULXs hosted by ETGs are considered less prone to variability, compared to ULXs hosted by late-type galaxies which tend to undergo more extreme inter-observational variability \citep{Bernadich2022}.    \par
The ULX candidate S5 stands out because of identifiable optical and FUV counterparts which is very rare for a ULX in a giant elliptical galaxy.  Point-like, compact optical emission in a galaxy can  be a knot of star formation or a stellar cluster or a nebula possibly related to supernova remnant and/or photoionized by the ULX similar to Holmberg II X-I \citep{Dewangan2004}.  Star formation knots (which can be bright in UV due to emission from hot OB stars) are not expected in  elliptical galaxies; therefore, the optical emission might be from a globular cluster.  Then, the FUV emission could be from hot HB stars in the globular cluster \citep{Peacock2017}. The optical emission can be from a nebula surrounding the ULX \citep{Pakull2003}. The FUV emission could be indicative of a strong accretion disk system also.  To understand more, a detailed study of the UV spectrum is needed.  S5 could be a HMXB because of the FUV counterpart; the spectrum is well fitted with an absorbed powerlaw augmented with bbody model for majority of the observations.  For S5, the average black body temperature is 0.15$\pm 0.04$ keV and average $\Gamma$ is $1.60\pm 0.31$.  The powerlaw component may have resulted from  Comptonization in a hot corona or accretion column over a NS surface. The remaining ULXs in NGC~5813 (S4, S6, S7, and S24) do not show any such counterparts in DSS or GALEX FUV images.  The best fit model for S4, S6 and S24 is absorbed Bbody, while for S7 it is absorbed Powerlaw. \par
The emitting radius of a source indicates the size of the region emitting X-rays, which in these cases is usually the accretion disk.   The emitting radius of S5 is of the order of 7,000 km which is around 30 times that of S4, S6 and S24, whose emitting radii lie in the range 190--250 km.  This could be either because the emitting surface is the photosphere of an extended outflow or because the compact object is an IMBH which will have a comparatively larger accretion disk. \par

Table~\ref{tab:my_table_5} gives the best-fit models and average luminosities of all the ULXs.  S6 has an average $\Gamma$ less than 1. It is the hardest among  these ULXs.  The spectra of ULX pulsars are the hardest among the ULXs \citep[e.g.][]{Pintore2017, Bachetti2020,Gurpide2021,Amato2023}.   So, S6 might be a possible candidate ULX pulsar, and may be hosting a neutron star \citep[e.g.][]{Carpano2018}.  S6 shows shifts in position in colour space (Figure~\ref{fig:my_label_cc}).\par 
Sources with luminosities above 10$^{39}$ erg/s appear to be present in about 30\% of galaxies across all morphological types but, there is slightly higher incidence of ULXs (nearly 40\%) in elliptical galaxies \citep{Kovlakas2020}.  One ULX was detected in NGC 4697 by \citet{Sarazin2000, Sarazin2001}.  \citet{Blanton2001} has studied the S0 galaxy NGC 1553, which has strong diffuse emission, and detected 3 ULXs.  NGC 1399 hosts eight ULXs; one of them is variable at a confidence level of 90\% \citep{Angelini2001, Feng2006}.  Seven ULXs were detected in NGC 720 by \citet{Jeltema2003}.  NGC 1407 is reported to have six ULXs with luminosity exceeding 10$^{39}$ erg/s \citep{Zhang2004}.  \citet{Humphrey2004} had identified two ULX candidates within the D25 radius of NGC 1332 but with luminosities below 2.0$\times$10$^{39}$ erg/s and they may simply be considered as the high-L$_X$ tail of LMXBs.   NGC 1600 is a central elliptical galaxy in a group.  It hosts 21 X-ray point sources with luminosities above 2.0$\times$10$^{39}$ erg/s -- approximately $11\pm2 $ of those sources are expected to be unrelated foreground or background sources \citep{Sivakoff2004}.  The detection of five ULXs in NGC~5813 is not surprising compared to the detections in NGC 1407, NGC 1600, NGC 1399 and NGC 720.  The detections in NGC 1407 and NGC 720 are based on a single observation but detections in NGC 1600 and NGC 1399 are based on multiple observations.  The ULX candidates (luminosity above 1.0$\times$10$^{39}$ erg/s) in NGC~5813 reported here, are detected in all the available nine \textit{Chandra} observations.  \par
Since ULXs are generally considered to be HMXBs and hence to be found in young stellar populations, the presence of ULXs in elliptical galaxies, that usually contain older stars and LMXBs, was somewhat puzzling.  \citet{Swartz2004}, \citet{David2005} and \citet{Plotkin2014} state that bright LMXBs can be ULXs as well.  We could not find any optical or UV counterparts for the ULXs S7, S6, S24 and S4.  So, there is a possibility that these ULXs are LMXBs. \par
ULXs in ETGs can be hosted by globular clusters also.  Two of the ULXs in NGC 1399 are hosted by GCs \citep{Angelini2001, Feng2006}.  \citet{Dage2019} has identified five GCULXs in NGC 4472, two in NGC 4649, and three in NGC 1399 (of which one GCULX faded beyond detection by 2005).  M 87 hosts seven GCULXs \citep{Dage2020}.  NGC 1316 hosts three ULXs that are associated with GCs \citep{Dage2021}.  \citet{Thygesen2023} has identified two GCULXs in NGC 1600, two in NGC 708, one in NGC 890, one in NGC 1060 and three in NGC 7619.  For NGC~5813, the  association with GCs was checked (GC data from \citep{Kundu2001}) (Figure~\ref{fig:my_label_fig2}).  It may be noted that \citet{Kundu2001} concentrated on the HST - WFPC2 field of view (2.7') and only S7 is within this field.  The closest GC is at 0.1' distance from S7.  Also, S5 is coinciding with an optical point source from the DSS observation; with a possible FUV counterpart also present in the GALEX image, it could be a GCULX.
 \par
Another possibility, to explain the presence of ULXs in elliptical galaxies, is that galaxy mergers can induce star formation in elliptical galaxies and create young stellar populations, which eventually give rise to ULXs \citep{Raychaudhury2008}.  There is no evidence of any recent mergers in NGC~5813; still there are evidences of a past merger event.  There are both metal poor and metal rich GCs in NGC~5813, which results in a significant optical Colour gradient of the GC population \citep{Hargis2014}.  The GCs have a significant age spread also \citep{Hempel2007}.  The dynamically de-coupled core and dust filaments in the inner regions of NGC~5813 were noted as reminiscent of a merger by \citet{Kormendy1984}.  So, there is a possibility that merger event/s might have induced a star formation event earlier in this galaxy and the ULXs might be the end products of the high mass stars formed during that phase.  There are evidences for AGN feedback with regular outbursts at an interval of 10$^7$ years in NGC~5813 \citep{Randall2011}. This might have subsequently quenched the star formation in the galaxy.
 \par
 NGC~5813 has SFR less than 0.003 M$_{solar}$/yr \citep{McDonald2021} and $M_{H_{2}}$ < 0.49 x 10$^8$ M$_{solar}$ \citep{McDonald2021}.  This clearly shows that NGC~5813 is a red galaxy with no recent star formation and  little molecular gas.  The presence of  ULXs in ETGs is correlated more with the metallicity of the galaxy than with their SFR. Metallicity in NGC~5813 falls from [Fe/H]>0 in the central regions to [Fe/H] < $\sim$ -0.7 at radii > $\sim$ 60" \citep{Efstathiou1985}.  All the five ULXs are in the region with [Fe/H] < $\sim$ -0.7.  This is in agreement with the anti-correlation of ULXs with local metallicity of the galaxy \citep{Kovlakas2020}.  
 
\section*{Acknowledgements}
The authors are very grateful to the anonymous referee for valuable comments which significantly improved the manuscript.  The scientific results reported in this article are mostly based on data obtained from the \textit{Chandra} Data Archive.  This research has made use of software provided by the \textit{Chandra} X-ray Center (CXC) in the application packages CIAO, ChIPS, and Sherpa. 
We extend our gratitude to Arunav Kundu, Principal Investigator, Eureka Scientific Inc. for providing the Globular Cluster catalogue of NGC~5813. We would like to thank Scott W. Randall, Harvard-Smithsonian Center for Astrophysics for providing the metal abundance map and plasma temperature map of NGC~5813.
The scientific results are based partly on observations obtained with XMM-Newton, an ESA science mission with instruments and contributions directly funded by ESA Member States and NASA.
The library facilities, ASSC server facility and the hospitality of Inter-University Centre for Astronomy and Astrophysics, Pune helped in the successful completion of this work.
Rajalakshmi T. R. acknowledges support for this work
from DST-INSPIRE Fellowship grant, IF170751, under
Ministry of Science and Technology, India.

\section*{Data Availability}
We have made use of \textit{Chandra} data and CIAO software for this work which are publicly available  at \textit{Chandra} X-ray Centre's \textit{Chandra} Data Archive\footnote[8]{\url{https://cxc.cfa.harvard.edu/cda/}} and CIAO web page\footnote[9]{\url{https://cxc.cfa.harvard.edu/ciao/}}.  The XMM-Newton data used for this work are also publicly available at ESA’s XMM-Newton Science Archive\footnote[10]{\url{http://nxsa.esac.esa.int/nxsa-web/}}.


\bibliographystyle{mnras}



\appendix

\bsp	
\label{lastpage}
\end{document}


\maketitle

\begin{landscape}
\begin{table*}
    \centering
    \begin{tabular}{|l|l|l|l|l|l|l|l|l|l|l|}
    \hline
     
    \multirow{2}{*}{Source} & \multirow{1}{*}{Obs Id} &
      \multicolumn{3}{c|}{$\chi ^2/dof$} &
      \multicolumn{3}{c|}{L (10$^{+39}$erg/s)} &
      \multicolumn{3}{c|}{Parameter} \\
     & & Model 1 & Model 2 & Model 3 & Model 1 & Model 2 & Model 3 & $\Gamma$ & kT (keV) & T$_{in}$ (keV) \\
    \hline
        S7 & 5907 & \textbf{4.65/7}  & 9.76/7  & \textbf{5.66/7} &  \textbf{3.02$^{+0.72}_{-0.67}$} &  2.13$_{-0.60}^{+0.54}$ & \textbf{2.62$^{+0.83}_{-0.69}$} & 1.30$^{+0.41}_{-0.39}$ & 0.86$^{+0.21}_{-0.17}$ & 1.99$^{+2.43}_{-0.71}$\\
        S7 & 9517 & \textbf{17.63/12} & 37.84/12 & 25.32/12 & \textbf{2.91$^{+0.40}_{-0.40}$} & - & - & 1.80$^{+0.25}_{-0.24}$ & - & -  \\
        S7 & 12951 & \textbf{6.30/8} & 12.63/8 & 8.48/8 & \textbf{3.71$^{+0.56}_{-0.56}$} & 1.80$^{+0.28}_{-0.28}$ & 2.26$^{+0.35}_{-0.34}$ & 2.16$^{+0.39}_{-0.39}$ &   0.37$^{+0.07}_{-0.06}$ &  0.59$^{+0.20}_{-0.13}$   \\
        S7 & 12952 & \textbf{23.80/21} & 86.91/21 & 52.38/21 & \textbf{4.16$^{+0.42}_{-0.42}$} & - & -  & 2.07$^{+0.18}_{-0.17}$ & - & - \\
        S7 & 12953 & \textbf{6.76/7} & 21.58/7 & 14.00/7 & \textbf{5.69$^{+1.04}_{-0.96}$} & - & 3.45$^{+0.61}_{-0.59}$ & 2.30$^{+0.40}_{-0.38}$ & - & 0.64$^{+0.25}_{-0.19}$  \\
        S7 & 13246 & 10.99/7 & 11.92/7 & \textbf{10.72/7} & 3.03$^{+0.67}_{-0.65}$ & 1.58$^{+0.34}_{-0.34}$ & \textbf{1.96$^{+0.43}_{-0.42}$} & 2.04$^{+0.50}_{-0.50}$  & 0.43$^{+0.11}_{-0.09}$ &  0.72$^{+0.35}_{-0.19}$  \\
        S7 & 13247 & \textbf{3.79/6} & \textbf{4.36/6} & \textbf{2.75/6} &  \textbf{3.76$^{+0.80}_{-0.78}$} & \textbf{2.22$^{+0.51}_{-0.49}$} & \textbf{2.69$^{+0.66}_{-0.59}$} & 1.74$^{+0.39}_{-0.39}$ & 0.55$^{+0.12}_{-0.10}$   & 0.97$^{+0.26}_{-0.44}$  \\
        S7 & 13253 & \textbf{16.17/16} & 42.76/16 & 25.00/16 & \textbf{3.87$^{+0.45}_{-0.44}$} & - & 2.57$^{+0.34}_{-0.32}$ & 1.95$^{+0.21}_{-0.21}$   & - & 0.81$^{+0.18}_{-0.14}$  \\
        S7 & 13255 & \textbf{4.74/8} & \textbf{7.11/8} & \textbf{4.93/8} & \textbf{4.22$^{+0.99}_{-0.84}$} & \textbf{2.21$^{+0.49}_{-0.45}$} & \textbf{2.81$^{+0.83}_{-0.60}$} & 1.68$^{+0.44}_{-0.43}$   & 0.52$^{+0.12}_{-0.10}$ & 0.95$^{+0.53}_{-0.26}$ \\ 
        & & & & & & & & & & \\
        S5 & 5907 & \textbf{53.86/33} & 196.13/33 & 133.16/33 & \textbf{14.65$_{-1.20}^{+1.22}$} & - & - & 2.14$_{-0.14}^{+0.14}$ & - & -  \\
        S5 & 9517 & \textbf{32.82/34} & 210.64/34 & 108.83/34 & \textbf{12.50$_{-0.91}^{+0.91}$} & - & - & 2.00$_{-0.10}^{+0.10}$ & - & -  \\
        S5 & 12951 & \textbf{17.43/24} & 156.78/24 & 79.92/24 & \textbf{13.34$_{-1.07}^{+1.09}$} & - & - & 1.93$_{-0.12}^{+0.12}$ & - & -  \\
        S5 & 12952 & \textbf{51.43/51} & 348.01/51 & 199.55/51 & \textbf{14.72$_{-0.93}^{+0.94}$} & - & - & 2.08$_{-0.08}^{+0.08}$ & - & -  \\
        S5 & 12953 & \textbf{14.22/14} & 87.52/14 & 44.08/14 & \textbf{18.74$_{-1.72}^{+1.73}$} & - & - & 2.06$_{-0.16}^{+0.15}$ & - & -  \\
        S5 & 13246 & \textbf{13.47/10} & 47.01/10 & 25.39/10 & \textbf{9.97$_{-1.10}^{+1.15}$} & - & - & 1.94$_{-0.21}^{+0.22}$ & - & -  \\
        S5 & 13247 & \textbf{22.36/17} & 80.97/17 & 54.32/17 & \textbf{12.65$_{-1.32}^{+1.30}$} & - & - & 2.12$_{-0.22}^{+0.21}$ & - & -  \\
        S5 & 13253 & \textbf{43.30/45} & 322.11/45 & 171.01/45 & \textbf{15.52$_{-1.02}^{+1.03}$} & - & - & 1.96$_{-0.08}^{+0.08}$ & - & -  \\
        S5 & 13255 & \textbf{21.55/21} & 100.87/21 & 54.43/21 & \textbf{12.41$_{-1.18}^{+1.19}$} & - & - & 2.06$_{-0.16}^{+0.16}$ & - & -  \\
        & & & & & & & & & & \\
        S6 & 5907 & \textbf{8.64/5} & 8.87/5 & 11.21/5 & \textbf{8.37$_{-2.00}^{+2.39}$} & 5.02$^{+1.74}_{-1.27}$ & - & 0.50$^{+0.35}_{-0.36}$ & 1.09$^{+0.31}_{-0.20}$ & - \\
        S6 & 9517 & 12.50/8 & \textbf{7.50/8} & 11.38/8 & 6.76$^{+0.94}_{-0.90}$ & \textbf{4.85$^{+0.76}_{-0.71}$} & 6.31$^{+1.15}_{-1.03}$ & 0.73$^{+0.18}_{-0.18}$ & 1.07$^{+0.13}_{-0.11}$ & 5.72$^{+12.79}_{-1.89}$ \\
        S6 & 12951 & \textbf{7.12/8} & \textbf{1.61/8} & \textbf{5.09/8} & \textbf{6.16$^{+1.22}_{-1.09}$} & \textbf{3.76$^{+0.74}_{-0.66}$} & \textbf{5.09$^{+1.50}_{-1.13}$} & 0.95$^{+0.26}_{-0.25}$ & 0.84$^{+0.12}_{-0.10}$ & 2.52$^{+3.29}_{-0.83}$\\
        S6 & 12952 & 8.43/8 & \textbf{4.44/8} & \textbf{6.89/8} & 4.58$^{+0.72}_{-0.67}$ & \textbf{2.99$^{+0.50}_{-0.46}$} & \textbf{4.10$^{+0.93}_{-0.77}$} & 0.80$^{+0.19}_{-0.19}$ & 0.95$^{+0.12}_{-0.10}$ & 3.95$^{+34.40}_{-1.52}$\\
        S6 & 12953 & \textbf{5.85/5*} & 2.94/5* & 4.93/5* & \textbf{4.50$^{+1.66}_{-1.18}$} & 2.85$^{+1.03}_{-0.72}$ & 3.74$^{+2.03}_{-1.14}$ & 0.88$^{+0.42}_{-0.43}$ & 0.88$^{+0.23}_{-0.15}$ & 2.77$^{+2.97}_{-0.90}$ \\
        S6 & 13246 & \textbf{2.99/3} & \textbf{1.17/3} & \textbf{2.49/3} & \textbf{5.63$^{+1.49}_{-1.29}$} & \textbf{3.59$^{+1.02}_{-0.85}$}& \textbf{4.97$^{+1.96}_{-1.44}$} & 0.86$^{+0.38}_{-0.38}$ & 0.93$^{+0.22}_{-0.16}$ & 3.46$^{+5.88}_{-1.12}$  \\
        S6 & 13247 & \textbf{1.54/4} & 8.49/4 & 2.82/5 & 5.29$^{+1.41}_{-1.27}$ &-& 4.55$^{+0.88}_{-0.88}$ & 0.81$^{+0.41}_{-0.40}$ &-& 3.4$^f$ \\
        S6 & 13253 & 15.98/10 & \textbf{8.70/10} & 15.45/10 & 4.44$^{+0.73}_{-0.70}$ & \textbf{3.05$^{+0.56}_{-0.51}$} & 4.11$^{+0.82}_{-0.81}$ & 0.66$^{+0.20}_{-0.20}$ & 1.03$^{+0.15}_{-0.12}$ & 7.01$^{+}_{-}$\\
        S6 & 13255 & 3.28/3 & \textbf{0.65/3} & \textbf{2.38/3} & 5.43$^{+1.41}_{-1.22}$ & \textbf{3.48$^{+0.93}_{-0.80}$} & \textbf{4.59$^{+1.83}_{-1.27}$} & 0.98$^{+0.38}_{-0.37}$ & 0.87$^{+0.19}_{-0.14}$ & 2.57$^{+45.95}_{-1.05}$ \\
        & & & & & & & & & & \\
        S24 & 5907 & 12.38/5 & \textbf{5.84/5} & 8.05/5 & - & \textbf{1.64$^{+0.33}_{-0.32}$} & 1.87$^{+0.40}_{-0.37}$ & - & 0.53$^{+0.08}_{-0.07}$ & 0.93$^{+0.28}_{-0.19}$\\
        S24 & 9517 & \textbf{10.21/11} & 26.67/11 & 12.31/11 & \textbf{4.12$^{+0.61}_{-0.56}$} & - & 3.01$^{+0.56}_{-0.48}$ & 1.55$^{+0.23}_{-0.22}$ & - & 1.17$^{+0.33}_{-0.23}$\\
        S24 & 12951 & 7.84/5 & 7.94/5 & \textbf{6.96/5} & 2.49$^{+0.57}_{-0.54}$ & 1.69$^{+0.47}_{-0.41}$ & \textbf{2.25$^{+0.66}_{-0.58}$} & 0.94$^{+0.30}_{-0.30}$ & 0.87$^{+0.21}_{-0.16}$ & 3.00$^{+14.74}_{-1.23}$ \\
        S24 & 12952 & 9.91/8 & \textbf{8.44/8} & 10.25/8 & 2.41$^{+0.45}_{-0.41}$ & \textbf{1.33$^{+0.23}_{-0.22}$} & 1.96$^{+0.53}_{-0.42}$ & 1.16$^{+0.25}_{-0.25}$ & 0.68$^{+0.09}_{-0.08}$ & 1.92$^{+1.36}_{-0.56}$ \\
        S24 & 12953 & 6.26/4* & \textbf{5.08/4*} & 6.05/5* & 3.26$^{+2.01}_{-1.15}$ & \textbf{1.82$^{+1.13}_{-0.59}$} & 2.58$^{+0.76}_{-0.64}$ & 0.77$^{+0.60}_{-0.61}$ & 0.86$^{+0.38}_{-0.20}$ & 3.01$^f$\\
        S24 & 13246 & 2.20/4* & \textbf{4.51/4*} & 2.65/4* & 2.19$^{+0.96}_{-0.55}$ & \textbf{1.18$^{+0.42}_{-0.29}$} & 1.51$^{+1.02}_{-0.42}$ & 1.55$^{+0.60}_{-0.59}$ & 0.56$^{+0.17}_{-0.11}$ & 1.11$^{+1.51}_{-0.38}$\\
        S24 & 13247 & \textbf{8.07/8*} & 6.85/8* & 7.32/8* & \textbf{2.76$^{+1.37}_{0.84}$} & 1.58$^{+0.70}_{-0.45}$ & 2.10$^{+1.76}_{-0.72}$ & 1.06$^{+0.52}_{-0.53}$ & 0.74$^{+0.23}_{-0.15}$ & 1.90$^{+1.25}_{-0.57}$\\
        S24 & 13253 & \textbf{6.72/7} & 9.18/7 & \textbf{5.12/7} & \textbf{2.59$^{+0.47}_{-0.44}$} & 1.64$^{+0.33}_{-0.30}$ & \textbf{2.15$^{+0.51}_{-0.44}$} & 1.14$^{+0.24}_{-0.24}$ & 0.77$^{+0.13}_{-0.11}$ & 1.97$^{+1.17}_{-0.54}$\\
        S24 & 13255 & 5.36/3* & \textbf{4.18/3*} & 4.73/3* & 2.49$^{+1.20}_{-0.71}$ & \textbf{1.41$^{+0.61}_{-0.38}$} & 1.87$^{+1.55}_{-0.60}$ & 1.16$^{+0.59}_{-0.58}$ & 0.72$^{+0.23}_{-0.15}$ & 1.71$^{+1.01}_{-0.51}$ \\
        & & & & & & & & & &\\

        \hline 
    
    \end{tabular}
    
\end{table*}
\begin{table*}
    \centering
    \begin{tabular}{|l|l|l|l|l|l|l|l|l|l|l|}
    \hline
    S4 & 5907 & 4.43/3* & \textbf{5.95/3*} & 7.04/4* & 4.49$^{+5.93}_{-2.19}$ & \textbf{2.11$^{+5.64}_{-0.99}$} & 1.81$^{+0.48}_{-0.42}$ & 0.47$^{+0.86}_{-0.87}$ & 0.96$^{+1.57}_{-0.34}$ & 1.5$^f$ \\
    S4 & 9517 & 7.21/4 & \textbf{3.48/4} & 4.94/4 & 1.98$^{+0.43}_{-0.39}$ & \textbf{1.24$^{+0.27}_{-0.25}$} & 1.55$^{+0.43}_{-0.35}$ & 1.28$^{+0.34}_{-0.32}$ & 0.71$^{+0.13}_{-0.10}$ & 1.57$^{+0.96}_{-0.44}$ \\
    S4 & 12951 & \textbf{2.64/3} & \textbf{2.22/3} & \textbf{2.17/3} & \textbf{1.99$^{+0.69}_{-0.52}$} & \textbf{1.09$^{+0.36}_{-0.28}$} & \textbf{1.45$^{+0.84}_{-0.44}$} & 1.30$^{+0.55}_{-0.53}$ & 0.66$^{+0.20}_{-0.14}$ & 1.45$^{+2.81}_{-0.56}$ \\
    S4 & 12952 & 8.46/8 & \textbf{7.81/8} & \textbf{5.36/8} & 2.04$^{+0.33}_{-0.31}$ & \textbf{1.33$^{+0.24}_{-0.23}$} & \textbf{1.60$^{+0.31}_{-0.28}$} & 1.51$^{+0.27}_{-0.25}$ & 0.69$^{+0.09}_{-0.08}$ & 1.33$^{+0.41}_{-0.27}$ \\
    S4 & 12953 & \textbf{7.40/7*} & 4.54/7* & 6.00/7* & \textbf{2.31$^{+0.96}_{-0.68}$} & 1.61$^{+0.71}_{-0.50}$ & 1.91$^{+1.01}_{-0.62}$ & 1.11$^{+0.52}_{-0.51}$ & 0.84$^{+0.27}_{-0.18}$ & 2.00$^{+7.25}_{-0.82}$ \\
    S4 & 13246 & 3.08/3 & \textbf{0.72/3} & \textbf{1.95/3} & 3.60$^{+0.98}_{-0.85}$ & \textbf{2.25$^{+0.62}_{-0.55}$} & \textbf{2.84$^{+1.10}_{-0.78}$} & 1.23$^{+0.47}_{-0.44}$  & 0.76$^{+0.18}_{-0.13}$ & 1.72$^{+2.24}_{-0.60}$ \\
    S4 & 13247 & \textbf{2.99/3*} & 1.35/3* & 2.24/3* & \textbf{2.90$^{+1.74}_{-0.91}$} & 1.47$^{+0.69}_{-0.41}$ & 1.99$^{+2.24}_{-0.67}$ & 1.19$^{+0.64}_{-0.63}$ & 0.63$^{+0.23}_{-0.14}$ & 1.45$^{+0.75}_{-0.43}$ \\
    S4 & 13253 & 7.41/5 & \textbf{2.98/5} & \textbf{4.80/5} & 2.18$^{+0.46}_{-0.40}$ & \textbf{1.26$^{+0.25}_{-0.23}$} & \textbf{1.59$^{+0.43}_{-0.33}$} & 1.39$^{+0.35}_{-0.33}$ & 0.64$^{+0.11}_{-0.09}$ & 1.32$^{+0.66}_{-0.34}$ \\
    S4 & 13255 & \textbf{2.21/4*} & 0.48/4* & 0.83/4* & \textbf{2.20$^{+0.73}_{-0.55}$} & 1.50$^{+0.53}_{-0.40}$ & 1.79$^{+0.76}_{-0.50}$ & 1.29$^{+0.47}_{-0.46}$ & 0.77 
    $^{+0.20}_{-0.14}$ & 1.68$^{+1.96}_{-0.57}$ \\
    & & & & & & & & & & \\
   
    \hline   
    \multicolumn{7}{|c|}{Model: Tbabs*(Apec+Bbody+Powerlaw)}  & & & & \\
    
    Source & Obs Id & $\chi ^2/dof$ & kT (keV) & $\Gamma$ & L (10$^{+39}$) & & & & & \\
    \hline
    S5 & 5907 & 42.51/31 & 0.10$^{+0.03}_{-0.04}$ & 1.76$^{+0.25}_{-0.27}$ & 17.03$^{+1.92}_{-1.82}$ & & & & & \\
    S5 & 9517 & 30.88/33 & 0.10$^f$ & 1.90$^{+0.15}_{-0.14}$ & 13.10$^{+1.20}_{-1.16}$ & & & & & \\
    S5 & 12951 & 12.63/22 & 0.13$^{+0.05}_{-0.05}$ & 1.68$^{+0.23}_{-0.27}$ & 14.41$^{+1.50}_{-1.44}$ & & & & & \\
    S5 & 12952 & 48.28/49 & 0.15$^{+0.05}_{-0.07}$ & 1.92$^{+0.17}_{-0.19}$ & 15.09$^{+1.12}_{-1.09}$ & & & & & \\
    S5 & 12953 & 13.33/12 & 0.13$^{+0.06}_{-0.08}$ & 1.93$^{+0.15}_{-0.14}$ & 19.44$^{+1.76}_{-1.76}$ $^a$ & & & & & \\
    S5 & 13246 & 12.02/9 & 0.13$^f$ & 1.71$^{+0.38}_{-0.35}$ & 10.87$^{+1.99}_{-1.67}$ & & & & & \\
    S5 & 13247 & 17.79/16 & 0.13$^f$ & 1.61$^{+0.42}_{-0.37}$& 14.74$^{+2.62}_{-2.24}$ & & & & & \\
    S5 & 13253 & 38.28/43 & 0.14$^{+0.05}_{-0.06}$ & 1.81$^{+0.15}_{-0.16}$ & 16.08$^{+1.23}_{-1.20}$ & & & & & \\
    S5 & 13255 & 20.56/20 & 0.14$^f$ & 1.90$^{+0.29}_{-0.27}$ & 12.79$^{+1.45}_{-1.36}$ & & & & & \\
    \hline    
    \end{tabular}
    \caption{Model comparison of candidate ULXs using \textit{Chandra} observations.     
     Model 1: Tbabs*(Apec+Powerlaw),\\
     Model 2:Tbabs*(Apec+Bbody), and
     Model 3: Tbabs*(Apec+Diskbb).  If, for a model the reduced chi$^2$ is greater than two, the luminosity and model parameters are not given in the table. $^a$ error obtained by freezing other parameters. $^f$ parameter is frozen to the value obtained for the closest observation. $^*$ \textit{C-stat / d o f} instead of $\chi ^2/dof$. $\chi ^2/dof$ and luminosity for the best fit model is given in bold} 
     \label{tab:tab2a}

\end{table*}

\end{landscape}